\documentclass[a4paper,twoside]{article}
\usepackage[dvips]{graphicx}
\usepackage{latexsym,epsf}
\usepackage{multicol}
\usepackage[centertags]{amsmath}
\usepackage [cp1251]{inputenc}
\usepackage [ukrainian, russian, english]{babel}
\usepackage {fancyhdr}

\renewcommand {\footrule}{\vbox to 0pt{\hbox to \headwidth{ \hrulefill \hspace{63mm}}\vss}}

\makeatletter
\renewcommand{\ps@plain}{
\renewcommand{\@oddhead}{}
\renewcommand{\@evenhead}{}
\renewcommand{\@oddfoot}{\hfil \thepage}
\renewcommand{\@evenfoot}{\thepage \hfil \hfil}}
\makeatother 

\makeatletter
\renewcommand{\@biblabel}[1]{#1.\hfill}
\makeatother
\title{\textbf{\Large FACTORIZATION AND QCD ENHANCEMENTS IN THE COMPTON MECHANISM OF W AND Z BOSON HADROPRODUCTION}}
\author{\textbf{\textit{M.V. Bondarenco\footnote{\normalfont
E-mail address: bon@kipt.kharkov.ua} }}
\\
\emph{\small National Science Center ``Kharkov Institute of
Physics and Technology",  61108,  Kharkov, Ukraine}\\
{\small(Received October 31, 2011)}}
\begin{document}
\selectlanguage{english}
\date{}
\maketitle

\thispagestyle{fancy}
\begin{center}
\begin{minipage}{165mm}
{\small
It is argued that $W/Z$ boson production in ultra-relativistic $pp$
collisions in the fragmentation region, subject to a kinematic cut
on the boson transverse momentum $Q_\perp>Q_{\perp\min}$, with 1
GeV/c$\ll Q_{\perp\min}\ll M_{W/Z}$, must be dominated by the
Compton mechanism $qg\to q'W/Z$. We propose applications for boson
hadroproduction in this kinematics and formulate the factorization
theorem. The analysis of QCD enhancements indicates that Regge
behavior should manifest itself not only in the $s$ but also in
$Q_\perp$-dependence of the boson production differential
cross-section. }
\par \vspace{1ex}
  \\
\end{minipage}
\end{center}
\begin{multicols}{2}
\begin{center}
\textbf{\textsc{1. INTRODUCTION}}
\end{center}\label{sec:intro}

Resonant production of $W^{\pm}$ and $Z^0$ bosons at $hh$ colliders
\cite{experimWZ} has high probability and a clean experimental
signature when the boson decays to a lepton pair, since both leptons
have $p_T\sim M_{V}/2\sim40$ GeV ($V=W$ or $Z$), thus being well
separated from the hadronic underlying event. That makes the
electroweak (EW) boson production process a convenient quark-meter,
well suited for determination of quark momentum distributions in
hadrons, complementary to DIS \cite{MSTW}, and also serve as a
playground for new physics searches.

Due to the $W$ boson charge, measurement of its asymmetry is
convenient for probing valence quark distributions
\cite{Berger-halzen-Kim-Willenbrock,Wasymm}. As for $Z$-bosons, they
have nearly the same differential distribution, but are easier to
reconstruct from detection of 2 charged leptons, and serve as a
benchmark. Albeit the formidable EW boson mass does not afford
probing the smallest $x$ frontier for the given proton beam energy
(at LHC in the central rapidity region $x_{1,2}\sim
M_{V}/{\sqrt{s}}\sim10^{-2}$), but in the fragmentation region one
of the $x$ diminishes to $\sim 10^{-3}$, so in this kinematics the
heavy boson hadroproduction may serve as a probe of small-$x$
physics, as well.

Historically, it was suggested by Drell and Yan \cite{Drell-Yan}
that the leading contribution to inclusive hadroproduction of a
heavy gauge boson comes from $q\bar q$ annihilation, with $q$ and
$\bar q$ carried by different hadrons (Fig.~1a). This approximation
holds well at $p\bar p$ collisions and for not very high energies
($\sqrt s\leq \frac{M_V}{x_{\text{val}}}\sim 0.5$ TeV). However, in
$pp$ collisions and at modern collider energies (particularly at
LHC), antiquarks are by far less abundant than
gluons\footnote{Gluons were not firmly established as partons at the
time article \cite{Drell-Yan} was written, whereas it had already
been known from electron DIS at SLAC that charged partons are
spin-$\frac12$ fermions, and so existence of quarks and antiquarks
was naturally hypothesized. Besides that, \cite{Drell-Yan} actually
discussed the case of finite $x$ for both fusing partons.}, and
since an EW boson can only be emitted from a quark line, the next
mechanism to be considered is $qg\to qV$ (known as QCD Compton
scattering, see Fig.~1b). Next, since at small $x$ gluons tend to be
more abundant than even quarks, in the central rapidity region one
also has to take into account boson production trough $gg$ fusion.
But as long as boson-gluon coupling is not direct, proceeding
through an auxiliary (virtual or real) $q\bar q$ pair production
(see Figs. 1c and 1d), $gg$ processes appear rather as a background
for quark physics studies, and it is desirable either to arrange
conditions in which their contribution is minor, or apply additional
experimental selection criteria to exclude them. Most obviously, to
suppress $gg$ fusion, it should suffice to work in the fragmentation
region, where $q(x)\gg \alpha_s g(x)$. Besides that the process at
Fig.~1d gives 2 jets \cite{2jet}, while that of Fig.~1c, as well as
Fig.~1a -- a minimum bias event with only small
$Q_\perp\sim\Lambda\leq1$ GeV. To compare with, the contribution
from the Compton mechanism is typically 1-jet and broadly
distributed in the boson's transverse momentum $Q_\perp$, out to
$Q_\perp\sim M_{V}$ (which at inclusive treatment of $Q_\perp$ gives
rise to $\ln\frac{M_{V}}{\Lambda}$). So, the criterion may be to
select events with 1 jet balancing the boson transverse momentum, in
favor of 2-jet and 0-jet events.

Strictly speaking, the physical distinction between the Drell-Yan
(DY) and Compton mechanisms is not quite clear-cut, because one of
the two Feynman diagrams of the Compton process (Fig.~2b) is
topologically similar to that of $q\bar q$ pair production by a
qluon and subsequent annihilation of the $\bar q$ with the
projectile quark. If in a proton all antiquarks stem from gluon
splitting $g\to q\bar q$, then one may expect the DY mechanism to be
contained in the Compton one. However, before the annihilation, the
antiquark may interact with other constituents of the hadron and get
non-trivially entangled with them, while in the Compton mechanism
this possibility is not accounted for. At evaluation of the
$Q_\perp$-integrated cross-section of boson production, that problem
is circumvented by taking the approach similar to DIS -- including
the LO gluon pdf contribution into NLO antiquark pdf contribution at
a higher factorization scale (actually $\sim M^2_{V}$)
\cite{Politzer,Altarelli}.

At high $p_T$, nonetheless, one may be pretty sure that most of the
(anti)quarks are due to direct gluon splitting. Although even an
intrinsic, low-$p_T$ antiquark can scatter on another constituent
(of the same hadron \cite{Feynman-Fox-Field}, or even of the
opposite hadron \cite{Berger-Brodsky}) and acquire high $p_T$, while
staying non-trivially entangled with the hadron constituents, but
the probability of hard scattering is $\sim\alpha^2_s(p^2_T)$,
whereas that of direct production of $\bar q$ at high $p_T$ from
gluon splitting is $\sim\alpha_s(p^2_T)$, i.e. of lower order in
$\alpha_s$ and thus greater. A caveat is that there are several
constituents in the proton to hard-scatter from, but from DIS
experience, that should not make the proton obscure with respect to
hard scattering, anyway. Besides that, the event shapes in these
cases contain different number of minijets and can be rejected by
additional experimental criteria. Therefore, at high $p_T$ the
distinction between DY and Compton mechanisms is clearer.

\begin{minipage}{75 mm}
\includegraphics{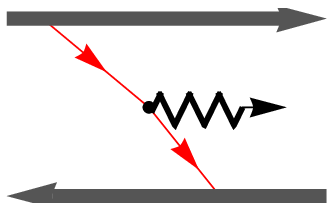}
\includegraphics{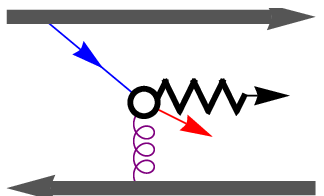}\\
$.\,\,\,\,\,\,\,\,\,\, \qquad \quad a) \qquad \qquad \qquad \qquad \qquad b)$ \\
\includegraphics{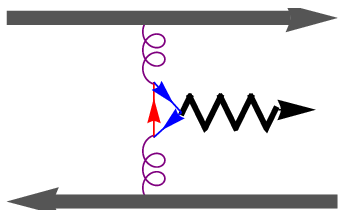}
\includegraphics{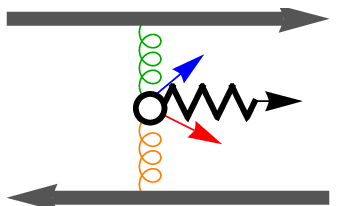}
$.\,\,\,\,\,\,\,\,\,\, \qquad \quad c) \qquad \qquad \qquad \qquad
\qquad d)$ \\
\emph{\textbf{Fig.1.}} {\emph{Contributions to $W$ and
$Z$ boson production in a high-energy proton-proton collision. (a)
-- $q\bar q$ annihilation; (b) -- $qg\to qV$ (Compton); (c) -- $gg$
fusion with virtual $q\bar q$ pair creation; (d) -- $gg$ fusion with
real
$q\bar q$ pair creation.}}\\
\end{minipage}

The above discussion suggests a possibility to experimentally
isolate the Compton contribution by imposing a kinematic cut
$Q_\perp>Q_{\perp\min}\gg \Lambda$, which suppresses the DY
contribution, and working in the fragmentation region, which
suppresses the contribution from $gg$-fusion. The contribution from
the central rapidity region may otherwise be suppressed by
constructing a $d\sigma(W^+)-\sigma(W^-)$ difference. The stipulated
high $Q_\perp$ will also alleviate the partonic subprocess
description, getting rid of collinear gluon reabsorption by the
non-annihilated active quark, and of the active quark irreducible
rescatterings \cite{Berger-Brodsky} (they are entirely absorbable
into $g(x_g)$). Hence, it may be feasible to use this mechanism as
well in global analysis of quark momentum distribution functions,
provided the factorization procedure is appropriately formulated. In
fact, the factorization procedure here must be similar to that in
case of direct photon or jet production, with the proviso that we do
not actually need to be involved in precise jet definition. That is
rather natural since the final quark with high $p_T$ will emerge as
a jet. Incidentally, let us note that tagging the flavor of the
quark jet, e.g. its charm, in association with a $W$ boson will give
access to the (sea) strangeness content of the nucleon (cf
\cite{Berger-halzen-Kim-Willenbrock}).

Comparing the procedures of pdf determination from EW boson
hadroproduction with a $Q_\perp$ cut and via inclusive $Q_\perp$
treatment, we should note the following. At LO, the DY mechanism is
very convenient for pdf determination because from the reconstructed
$V$-boson momentum one determines longitudinal momenta of both
annihilating partons exactly. At NLO, though, due to an additional
unregistered gluon in the final state it involves an
$x$-convolution. For Compton mechanism, the convolution arises
already at LO. However, since diagram 2b is similar to that of DY,
and moreover, it appears to dominate, the convolution kernel for
Compton is also rather singular, and perhaps even replaceable by a
$\delta$-function. This is quite opposite, say, to the direct photon
production case, where the emitted photon is always the lighter
particle among the final products, and therefore tends to carry away
only a small fraction of energy. For heavy boson production, the
roles of the momentum-conserving radiator and the soft radiation are
reversed: the emitted boson assumes most of the momentum while the
left-over quark is wee. The dynamical reason is that diagram 2b
dominates, wherein the final quark and the virtual space-like quark
tend to be collinear with the parent gluon, and so the space-like
quark manifests itself more like an antiquark.

\begin{minipage}{65 mm}
\includegraphics{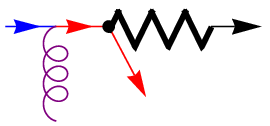}
\includegraphics{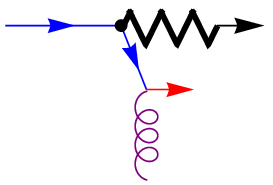}
$.\,\,\,\,\,\,\,\,\,\, \qquad  a) \qquad \qquad \qquad \qquad
\qquad b)$ \\
\emph{\textbf{Fig.2.}} {\emph{Feynman diagrams for the
partonic
subprocess of Fig.~1b (Compton scattering).}}\\
\end{minipage}

Ultimately, the dominance of diagram 2b may be utilized for probing
the gluon distribution in a close analogy with probing quark
distribution at DIS. Indeed, the virtual space-like ($\sim-M^2_V$)
quark in Fig.~2b corresponds to the virtual photon in the DIS LO
diagram, and this virtual quark knocks out a quasi-real gluon from
the proton, converting it to a quark. However, at small $x_g$ one
must beware of multiple gluon exchanges between the probing quark
and the probed hadron, which can affect the universality of the
probabilistic gluon distribution. Those issues will be discussed in
the next section.

\begin{center}
\textbf{\textsc{2. FACTORIZATION FOR THE COMPTON PRODUCTION
MECHANISM BEYOND GLUON DISTRIBUTIONS}}
\end{center}

In this section we outline the factorization procedure relating the
fully differential cross-section of EW boson hadroproduction with
the corresponding $qg\to q'V$ partonic cross-section. At small $x$
values, it may be important to formulate the factorization theorem
non-perturbatively, beyong the notion of gluon distribution
probability. The irreducibility to single gluon exchange amounts to
quark scattering off an intense and coherent gluonic field. In this
respect the situation resembles that in QED, where there is a
familiar factorization theorem beyond the perturbative treatment of
interaction with the external field, first established for
scattering in a Coulomb field \cite{OMW}, and then generalized to
high-energy scattering in compact field of an arbitrary shape
\cite{Baier-Katkov} (see also \cite{Bond}). The theorem presents the
process amplitude as a product of the electron spin-independent
(eikonal) scattering amplitude and the electron spin-dependent
perturbative amplitude of real photon emission at absorption of a
virtual photon with the momentum equal to the total momentum
transfer in scattering. In our case the initial gluon virtuality may
be neglected compared with the emitted boson mass. Therefore we may
regard the initial gluon as real, avoiding the issue of initial
gluon distribution with the account of virtuality, and apply the
Weizs\"{a}cker-Williams approximation. The latter, however, needs to
be generalized, factoring out not the gluon flow but the full quark
scattering amplitude.

Consider a non-diffractive high-energy $pp$ collision event
containing a high-$p_T$ $l\bar l$ pair. Suppose that by
reconstructing the total momentum of the $l\bar l$ pair its mass is
identified to be at the $W$ or $Z$ boson resonance\footnote{The
resonance width $\Gamma\sim2 \text{GeV}\ll M_{V}$ will be neglected
in this article throughout, and so the boson is handled as a
quasi-free particle.}, and the rapidity being $>1$, say, positive.
The latter implies that this boson had most probably been emitted by
one of the quarks of the hadron moving in the positive (forward)
direction.

Owing to the Lorentz-contraction of ultra-relativistic hadrons, the
interaction of the emitter quark with the opposite hadron proceeds
very rapidly. Furthermore, owing to large value of $M_V$ compared to
the typical hadronic energy scale 1 GeV, the boson emission from the
quarks also passes very rapidly compared to the intra-hadron
timescale. Hence, sufficiently reliable must be the impulse
approximation, at which the emitting quark initial state is
described by the (empirical) momentum distribution function, while
the rest of the partons in that hadron are regarded as spectators
taking no part in the boson production process. Thereby we reduce
the problem to that of $V$ boson emission by a relativistic quark
scattering on an hadron. Since due to the boson heavyness, the
amplitudes of its emission from different quarks within one (forward
moving) hadron do not interfere (the formfactor reduces to the
number of quarks), the probability (differential cross-section) of
boson production in the $pp$ collision comes as an integral of the
correspondent quark-proton differential cross-section weighted with
the quark pdf $f(x)$ in the first proton:
\begin{equation}\label{quark-pdf}
d\Sigma(P_1,P_2,Q)=\int_0^1\! dx f(x) d \sigma(xP_1,P_2,Q)\quad
(\text{fr. reg.}),
\end{equation}
where $P^{\mu}_1$, $P^{\mu}_2$ are the initial hadron 4-momenta, $Q$
the final boson momentum, and $x$ the hadron momentum fraction
carried by the emitter quark. The factorization scale for the quark
pdf $f(x)$ will be determined later on.

\textbf{{Relation with the quark-hadron scattering
differential cross-section}}\\

Applying the generalized Weizs\"{a}cker-Williams procedure to the
differential cross-section of boson production in quark-hadron
scattering, we obtain
\begin{equation}\label{16}
    \frac{d\sigma}{d\Gamma_Q}=16\pi\frac{2E}{E'+p'_z}\frac{d\hat{\sigma}}{dt}\mathbf{k}^2_{\perp}d\sigma_{\text{scat}},
\end{equation}
where
\[
    d\Gamma_Q=\frac{d^3Q}{(2\pi)^32Q_0},
\]
and $d\hat{\sigma}$ may be related with QED of QCD virtual Compton
cross-section:
\[
    \frac{d\hat{\sigma}}{dt}=\frac1{4\pi\alpha_{\text{em}}}\frac{d\sigma(e\gamma\to eV)}{dt}
    =\frac{2N_c}{4\pi\alpha_s}\frac{d\sigma(qg\to qV)}{dt}.
\]
The value of the gluon longitudinal momentum, or energy, $\omega$ is
fixed by the 4-momentum conservation law and on-shellness of the
final undetected quark.

The quark-hadron quasi-elastic scattering differential cros-section
$d\sigma_{\text{scat}}$ encodes all non-perturbative aspects of fast
quark-hadron interaction in a model-independent way. Earlier, the
differential cross-section of quark-hadron scattering had already
been introduced in the context of high-energy $pA$ \cite{BDMPS} and
$\gamma A$, $\gamma^* A$ collisions of nucleons and nuclei.

Representation (\ref{16}) also expresses similarity with
$k_T$-factorization \cite{kT-fact}, but there $d\hat\sigma$ may go
beyond the WW approximation, while in place of
$d\sigma_{\text{scat}}$ one has the BFKL kernel describing the
growth of the cross-section with the energy. In the next section we
shall discuss the latter issue as well, along with other effects
arising in QFT.

\textbf{Relation with the integrated and unintegrated gluon distributions}\\

If we could rely on an approximation that the high-energy small
angle quark-hadron scattering proceeds only through a single
$t$-channel gluon exchange (presumably with a running coupling
constant), we might avoid detailed description of the hadron
creating the color field. Then all we need to know is the equivalent
gluon flow. In the single gluon exchange approximation, the absorbed
gluon may be treated on equal footing with the initial quark, and so
the description of boson hadroproduction must become symmetric in
terms of initial quark and gluon distributions.

Implementing the single gluon exchange approximation into the
factorization procedure and comparing the final result with
Eq.~(\ref{16}), we obtain a formula for the unintegrated gluon
density (DGLAP type)
\begin{equation}\label{unint}
        x_gg(x_g,k^2_\perp,Q^2_\perp)=\frac1\pi\frac{2N_c}{4\pi\alpha_s(Q^2_\perp)}
    \mathbf{k}^2_\perp\frac{d\sigma_{\text{scat}}}{d^2k_\perp}.
\end{equation}
This function vanishes at $k_\perp\to0$ due to factor
$\mathbf{k}^2_\perp$, as well as at $k_\perp\to\infty$ due to factor
${d\sigma_{\text{scat}}}/{dk^2_\perp}$. Hence, somewhere in between
it must have a maximum, but it is unobvious whether it belongs to
the hard or soft region, and whether the decrease immediately beyond
the maximum follows exponential or power law
$\sim1/\mathbf{k}^2_\perp$. The existing parameterizations favor the
hard scenario.

The corresponding conventional gluon density obtained by
$k_\perp$-integration of Eq.~(\ref{unint}) is
\begin{equation}\label{g-int}
    x_gg(x_g,Q^2_\perp)=\frac1\pi\frac{2N_c}{4\pi\alpha_s(Q_\perp)}\int_0^{Q^2_\perp}dk^2_\perp
    k^2_\perp\frac{d\sigma_{\text{scat}}}{dk^2_\perp}.
\end{equation}
This may be compared with DIS in the dipole picture
\cite{dipole-DIS}, and with the approach of
\cite{Kimber-Martin-Ryskin}.

\begin{center}
\textbf{\textsc{3. MODIFICATIONS ARISING IN QFT}}
\end{center}

In ordinary quantum mechanics, the differential cross-section of
$qh$ scattering appearing in Eqs.~(\ref{16}, \ref{unint}) whould
assume a finite value in the high-energy limit. But it is now
well-known that QFT brings (fortunately, mild) modifications to the
impulse approximation and the parton model, for a number of reasons.
First of all, even a static field created by an ensemble of
point-like partons has Coulomb singularities, resulting in a
logarithmic dependence of the transport or radiative cross-section
on some hard scale. Secondly, multiple emission of soft quanta and
particle pairs in the central rapidity region generates various
double logarithmic asymptotics in the cross-ection, which upon
resummation to all orders may turn into power-law modifications
\cite{Gorshkov}. We shall discuss these effects by turn as applied
to our specific problem.

\textbf{$Q_\perp$ as the natural factorization scale for Compton
mechanism}

At practice, transverse momenta of multiple final hadrons produced
within the underlying event are usually not counted, and
correspondingly, $d\sigma/d\Gamma_Q$ must be integrated over the
unconstrained momentum components of the initial gluon(s) -- i.e.,
over $\mathbf{k}_\perp$. In so doing, it seems reasonable to neglect
the $\mathbf{k}_\perp$-dependence of the Compton subprocess
cross-section $d\hat{\sigma}/dt$ provided $\mathbf{k}_\perp^2\ll
p\cdot k$. That leads to
\begin{equation}\label{dsigma-int}
    \frac{d\sigma}{d\Gamma_Q}=16\pi\frac{2E}{E'+p'_z}\frac{d\hat{\sigma}}{dt}\int
    d^2k_\perp\mathbf{k}^2_{\perp}\frac{d\sigma_{\text{scat}}}{d^2k_\perp}.
\end{equation}

However, at large $\mathbf{k}^2_{\perp}$ the scattering differential
cross-section has Rutherford asymptotics (cf., e.g.,
\cite{semi-hard}):
\begin{equation}\label{sigma-asymp}
    \frac{d\sigma_{\text{scat}}}{d^2k_\perp}\underset{k_\perp\to\infty}\sim
    \frac{2\alpha^2_s(k^2_\perp)}{\mathbf{k}^4_{\perp}}\left[\left(1-\frac1{N^2_c}\right)\frac{N_q+N_{\bar q}}2+N_g \right]
\end{equation}
(with $N_c=3$ the number of colors, and $N_q$, $N_{\bar q}$, $N_g$
the mean numbers of quarks, antiquarks and gluons in the proton),
and therewith the $k_\perp$-integral in (\ref{dsigma-int}) appears
to be logarithmically divergent at the upper limit. That means that
at sufficiently large $k_\perp$ one still needs to rely on the
decrease of ${d\hat{\sigma}}/{dt}$ with $k_\perp$, providing the
additional convergence factor. The sensitivity of
${d\hat{\sigma}}/{dt}$ to $k_\perp$ arises at
\begin{equation}\label{kTmax}
    k_{\perp\max}\sim\min\{M_{V},Q_\perp\},
\end{equation}
which should be used as the upper limit in $k_\perp$ integral in
Eq.~(\ref{dsigma-int}) and serve as a natural factorization scale.
It is also to be used as a factorization scale for the quark pdf in
Eq.~(\ref{quark-pdf}), if we wish at determination of $Q_\perp$ to
be able to neglect the initial quark transverse momentum. In what
follows, we will be mostly considering the case $Q_\perp<M_{V}$,
whereby $k_{\perp\max}\sim Q_{\perp}$. That differs from the case of
DY mechanism, where even at small $Q_\perp$ the natural
factorization scale is $M_{V}$
\cite{Politzer,M-fact-scale}\footnote{For $Q_\perp$-integrated
distributions, the factorization scale is usually taken to be $\sim
M^2_V$, as well -- see \cite{Anastasiou}.}, and is in the spirit of
factorization in direct photon and jet production \cite{Owens}.

\textbf{Distribution of the color sources}

With asymptotics (\ref{sigma-asymp}) and upper limit (\ref{kTmax}),
the $k_\perp$-integral in (\ref{dsigma-int}) with constant $N_q$,
$N_g$ would give $\ln\frac{Q_\perp}{\Lambda}$. But in fact, $N_q$,
$N_g\neq \mathrm{const}$, because they express as integrals from
pdfs, which diverge at low $x$. If the asymptotics of $f(x')$ is
$\sim1/x'$, as is motivated by the perturbation theory
\begin{equation}\label{Nq}
    N_q=\int^1_{\mathbf{k}^2_\perp/xs}dx'f(x')\sim\ln\frac{x
    s}{\mathbf{k}^2_\perp},
\end{equation}
where $xs$ is the quark-hadron collision subenergy. Substituting
Eqs.~(\ref{sigma-asymp}, \ref{Nq}) to (\ref{dsigma-int}), we get
\begin{eqnarray}\label{}
    \int
    dk^2_\perp\mathbf{k}^2_{\perp}\frac{d\sigma_{\text{scat}}}{d^2k_\perp}
    \sim\int^{Q^2_\perp}_{\Lambda^2}d\ln\mathbf{k}^2_\perp
    \ln\frac{xs}{\mathbf{k}^2_\perp}\nonumber\\
    \simeq\frac12\ln^2\frac{x
    s}{\mathbf{k}^2_\perp}\Bigg|_{\mathbf{Q}^2_\perp}^{\Lambda^2}=\ln\frac{\mathbf{Q}^2_\perp}{\Lambda^2}\ln\frac{xs}{\Lambda
    |\mathbf{Q}_\perp|}.
\end{eqnarray}
This equation is similar to the (Sudakov) double logarithms for the
reggeized gluon, if $\mathbf{Q}^2_\perp$ stands for $|t|$. But
$\ln\frac{\mathbf{Q}^2_\perp}{\Lambda^2}$ may be absorbed into pdf
definition.

In a more empirical approach, however, the divergence proceeds as a
power law \cite{Barone-Predazzi}
\begin{equation}\label{}
    f(x')\sim \left(\frac1{x'}\right)^{\alpha_{\text{P}}}, \qquad
    \alpha_{\text{P}}>1,
\end{equation}
and the factorization scale must be taken $\ll Q_\perp$ (cf. CGC
approach \cite{CGC}). The $x'$-integration then gives
\begin{equation}\label{}
    N_q\sim\left(\frac{x
    s}{\mathbf{k}^2_\perp}\right)^{\Delta},\qquad
    \Delta=\alpha_{\text{P}}-1>0,
\end{equation}
Therewith, the $k_\perp$-integral in Eq.~(\ref{dsigma-int})
converges on the upper limit
\begin{equation}\label{}
    \int d^2k_\perp
    k^2_\perp\frac{d\sigma_{\text{scat}}}{dk^2_\perp}\sim (x
    s)^{\Delta}
    \int^\infty_{\Lambda^2}\frac{dk^2_\perp}{(k^2_\perp)^{\alpha_{\text{P}}}}
    \sim \left(\frac{x
    s}{\Lambda^2}\right)^{\Delta},
\end{equation}
and the result is independent of the factorization scale, provided
$\Delta$ is. That must correspond to the BFKL-regime
\cite{Barone-Predazzi}.

\textbf{Small-$x$ behavior of the gluon distribution}

Since at present we can not reliably calculate the gluon
distribution function \emph{ab initio}, it is to be inferred on
phenomenological basis. Although gluon is not directly observable
outside of $hh$ collisions, in DIS at small $x$ its density must be
proportional to that of sea quark, which, in turn, is $\propto
F_2(x,\mu^2)$. At typical scales of hadroproduction at LHC
($Q^2_\perp\sim100\, \text{GeV}^2$ and $x_g\sim 10^{-3}$), the DIS
data for the nucleon structure function are not available, but the
data come close, and seemingly admit safe extrapolation. One must
also duly incorporate the dependence on the factorization scale
$\mu^2$, since at so small $x$ it is significant. To extrapolate
both $x$ and $Q^2$ dependences, one may utilize the observation
\cite{geom-scaling} that at $x<0.01$ the DIS $\gamma^* p$
cross-section
\begin{equation}\label{}
    \sigma_{\gamma^* p}=\frac{4\pi^2\alpha_{\text{em}}}{\mu^2}F_2(x,\mu^2)=\sigma_{\gamma^*
    p}(\tau)
\end{equation}
obeys ``geometrical scaling", reducing to a function of a single
variable
\begin{equation}\label{tau}
    \tau=\frac{\mu^2}{\mu^2_0}\left(\frac{x_g}{x_0}\right)^\lambda,\quad \mu_0=1\,\text{GeV},
\end{equation}
with the best-fit parameters \cite{geom-scaling}
\[
    \lambda\approx0.3,\quad x_0\approx
    3\cdot 10^{-3}.
\]
Furthermore, in the domain $\tau\gg1$, to which our parameters
belong, the dependence on $\tau$ is a simple power law in itself:
\begin{subequations}
\begin{eqnarray}
    F_2(x,\mu^2)&=&\frac{\mu^2}{4\pi^2\alpha_{\text{em}}}40\mu\text{b}\,\tau^{-\Delta/\lambda}\label{34a}\\
    &=&0.35\left(\frac{\mu^2}{\mu^2_0}\right)^{1-{\Delta}/{\lambda}}\left(\frac{x_0}{x}\right)^{\Delta}.\label{34b}
\end{eqnarray}
\end{subequations}
Next we note that phenomenologically the exponent in Eq.~(\ref{34a})
$\Delta/\lambda\approx0.75$, and so in Eq.~(\ref{34b})
$1-\Delta/\lambda\approx0.25\approx\Delta$, i.e. exponents for
$\mu^2$ and $1/x$-dependencies are equal. Theoretically, there might
be some difference between them in connection that integral $\int
d^2k_\perp$ diverges and demands a cutoff at $\sim \mu^2$. But if
for simplicity we assume the equality of the exponents, and utilize
the relation
\begin{equation}\label{}
    \mu^2/x=W^2+\mu^2\approx W^2,
\end{equation}
\begin{equation}\label{29}
    \alpha_s(\mu^2)x_gg(x_g,\mu^2)\propto
F_2(x,\mu^2)\approx0.07\left(\frac{W^2}{\mu_0^2}\right)^{0.25}.
\end{equation}
Recalling the relation with the differential cross-section
(\ref{unint}), equation (\ref{29}) is quite natural from the
viewpoint of $t$-channel Reggeization. Then, we obtain the same
result in any treatment -- through pdfs or through $qh$ scattering.

From the viewpoint of DGLAP equations, however, the factorization
and Reggeization sooner or later must break down, and saturation to
set in. In that case, the slope of $Q_\perp$-dependence should vary
with the energy.

\textbf{Reggeization in the $Q_\perp$-dependence of the boson
hadroproduction differential cross-section}

In general, the onset of an energy-dependence of the quark-hadron
scattering amplitude, along with the strong difference between
initial and final quark energies in the hard subprocess may affect
the balance of Compton process Feynman diagrams, and in principle
violate the gauge invariance. Fortunately, at $Q_\perp\ll M_{V}$,
only one of the two Feynman diagrams dominates, wherein the final
quark interacts with the encountered proton. Then, the $qh$
collision subenergy is counted by the energy of the final quark. To
estimate it, note that
\begin{equation}\label{}
    x_g\sim M^2_V/xs,
\end{equation}
\begin{equation}\label{subenergy}
    p'\cdot P_2=\frac{p'\cdot k}{x_g}=\frac{p'\cdot k}{p\cdot
    k}xs=\frac{p'_+}{p_+}xs=\frac{\mathbf{Q}^2_\perp}{p\cdot p'}xs\sim
    \frac{\mathbf{Q}^2_\perp}{M^2_V}xs.
\end{equation}
The non-trivial thing about subenergy (\ref{subenergy}) is the
${\mathbf{Q}^2_\perp}/{M^2_V}=\rho^2$ factor. It means that as a
result of Reggeization, the differential cross-section multiplies by
$\rho^{2\Delta}$. That factor may also be considered as being due to
the factorization scale $Q_\perp$ dependence of the gluon structure
function. The rest of the $Q^2_\perp$-dependence comes from the
partonic Compton differential cross-section, which in the
perturbative description (the sum of diagrams 2a and 2b) goes as
$\sim Q^{-2}_\perp$ for $\Lambda\ll Q_\perp\ll M_V$ and as $\sim
Q^{-4}_\perp$ for $Q_\perp\geq M_V$. Hence, in the fragmentation
region of rapidities, and intermediate region of boson transverse
momenta $\Lambda\ll Q_\perp \ll M_V$ the boson hadroproduction
differential cross-section should behave as
\begin{equation}\label{QT-12}
    \frac{d\sigma}{dQ_\perp}\sim Q_\perp^{2\Delta-1}\sim
    Q_\perp^{-1/2} \qquad(\text{fragm. region}).
\end{equation}

In the central rapidity region, $Q_\perp$-dependence of the quark
pdf comes into play, and Eq.~(\ref{QT-12}) modifies to
\begin{equation}\label{}
    \frac{d\sigma}{dQ_\perp}\sim Q_\perp^{4\Delta-1}\sim
    Q_\perp^{0} \qquad(\text{centr. region}).
\end{equation}
However, one must keep in mind that in the central region there are
other contributing mechanisms besides the Compton one.

\textbf{Sudakov form-factors in the Compton subprocess}

\nopagebreak The above discussed power-law increase of sea and gluon
pdfs at small $x$ physically owes to the opening possibility of
particle production in the central rapidity region, with the phase
space indefinitely increasing with the collision energy.
Theoretically it is connected with gluon Reggeization and double
logarithmic asymptotics. But double log effects also arise in the
hard subprocss, since $M^2_{V}$ is a large scale, while $Q^2_\perp$
is a smaller subscale. Physically, at color exchange the particle
intensely emit soft and collinear radiation quants, but those do not
essentially change the emitting particle energy, only alter the
transverse momentum. For large-angle scattering, this is
inessential, but for small-angle it is. Resummation of real and
virtual contributions does not compensate completely, which leads to
transverse Sudakov form-factors.

The Sudakov resummation for the Drell-Yan production mechanism at
$Q_\perp\ll M_{V}$ received a great deal of attention (see, e.g.,
\cite{DY-resummation} for an overview). But for the Compton
mechanism the implementation of the developed techniques is hampered
by the presence of 3 eikonal lines instead of 2 in DY, though all
belonging to one plane.

The Sudakov formfactors also lead to a hardening of the
$Q_\perp$-dependence. But basically it is just a redistribution,
whereas Regge effects are pure enhancements. So far, Tevatron
\cite{Tevatron-QT} and LHC \cite{LHC-QT} agreed well with the
predictions of existing Monte-Carlo generators. Thus, in order to
discern Regge effects in the $Q_\perp$-dependence and disentangle
them from Sudakov effects, more studies are required.

\begin{center}
\textbf{\textsc{4. CONCLUSIONS}}
\end{center}

In the present note we have argued that the Compton mechanism of EW
boson hadroproduction, when properly isolated from the other
mechanisms, may serve for determination of quark distributions in
the proton in the valence domain. We have established the recipe for
the pdf factorization scale ($Q^2_\perp$ instead of $M^2_V$), and
formulated qualitative predictions for modification of
$Q_\perp$-dependence of the boson hadroproduction cross-section due
to the physical manifestation of the factorization scale, which can
also be viewed as manifestation of Reggeization in the quark-hadron
scattering.

In conclusion, let us point out the experimental objectives ensuing
from our study. In order to measure the pdfs in the proton it might
suffice to deal with boson rapidity distributions, but to control
the underlying partonic dynamics, and actually to test the
non-trivial predictions of QCD in hard but non-perturbative regime,
one also needs measurement of the $Q_\perp$-dependence. With the
accumulation of statistics at LHC, it is desirable to measure at
ATLAS and CMS the double-differential (in rapidity and $Q_\perp$)
cross-section of $Z$-boson production. In the central rapidity
region and in the fragmentation region the $Q_\perp$-dependencies
may differ. The same procedure is to be repeated for (absolute)
cross-sections of $W^+$ and $W^-$ boson production, since it is the
difference of those cross-sections which determines the valence
quark distribution function in the proton. Although the
reconstruction of $W$ boson transverse momentum is more complicated
because of unregistered neutrinos, at the Tevatron it was managed
\cite{W-QT-Tevatron}, and must be feasible at LHC.

\textbf{Acknowledgement}

Thanks are to P.V. Sorokin for inspiring my interest to the EW boson
production problems at LHC.

\vspace{3mm}
\begin{center}

\end{center}
\end{multicols}


\begin{thebibliography}{99}
\bibitem{experimWZ}
R.~Stroynowski. Phys. Rep. \textbf{71} (1981) 1; \\
J.~Alitti et al. Z. Phys. C \textbf{47} (1990) 11; \\
T.~Aaltonen et al., CDF Collaboration, Phys. Lett. B \textbf{692} (2010) 232; \\
V.M.~Abazov et al., D0 Collaboration, Phys. Lett. B \textbf{693}
(2010) 522; \\
V.~Khachatryan et al., CMS Collaboration. JHEP \textbf{01} (2011)
080.

\bibitem{MSTW}
P.M.~Nadolsky et al. Phys. Rev. D \textbf{78} (2008) 013004; \\
A.D.~Martin, W.J.~Stirling, R.S.~Thorne, and G.~Watt. Eur. Phys. J.
C \textbf{63} (2009) 189; \\
M.~Dittmar et al. Parton Distributions. e-Print: arXiv:0901.2504
[hep-ph]; \\
M.~Dittmar, F.~Pauss, and D.~Zurcher. Phys.Rev. D \textbf{56} (1997)
7284.

\bibitem{Berger-halzen-Kim-Willenbrock}
E.L.~Berger, F.~Halzen, C.S.~Kim, and S.~Willenbrock. Phys. Rev. D
\textbf{40} (1989) 83.

\bibitem{Wasymm}
T.~Aaltonen et al., CDF Collaboration, Phys. Rev. Lett. \textbf{102}
(2009) 181801; \\
G.~Aad et al., ATLAS Collaboration, Phys. Lett. B \textbf{701}
(2011) 31.

\bibitem{Drell-Yan} S.D.~Drell and T.M.~Yan. Phys. Rev. Lett. \textbf{25} (1970) 316
[Erratum-ibid. \textbf{25} (1970) 902].

\bibitem{2jet}
J.~Campbell and R. K. Ellis. Phys. Rev. D \textbf{65} (2002) 113007.

\bibitem{Politzer}
H.D.~Politzer. Nucl. Phys. B \textbf{129} (1977) 301; \\
C.T.~Sachrajda. Phys. Lett. B \textbf{73} (1978) 185.

\bibitem{Altarelli}
G.~Altarelli. Phys. Rep. \textbf{81} (1982) 1; \\
G.~Sterman et al. Rev. Mod. Phys. \textbf{67} (1995) 157.

\bibitem{Feynman-Fox-Field}
R.P.~Feynman, R.D.~Field, and G.C.~Fox. Phys. Rev. D \textbf{18}
(1978) 3320;  \\
R.D.~Field. Talk at 19th Int. Conf. on High Energy Physics, Tokyo,
1978.

\bibitem{Berger-Brodsky}
E.L.~Berger and S.~Brodsky. Phys. Rev. Lett. \textbf{42} (1979) 440.

\bibitem{OMW}
H.~Olsen, L.C.~Maximon, and H.~Wergeland, Phys. Rev. \textbf{106}
(1957) 27.

\bibitem{Baier-Katkov}
V.N.~Baier and V.M.~Katkov, Sov. Phys. JETP \textbf{26} (1968) 854;
\textbf{28} (1969) 807.

\bibitem{Bond}
M.V.~Bondarenco. Phys. Rev. A \textbf{82} (2010) 042723.

\bibitem{BDMPS}
E. Levin. Phys. Lett. B \textbf{380} (1996) 399; \\
R. Baier, Yu.L.~Dokshitzer, A.H.~Mueller, S.~Peigne, and D.~ Schiff.
Nucl. Phys. B \textbf{484} (1997) 265.

\bibitem{kT-fact}
S.~Catani, M.~Ciafaloni, and F.~Hautmann. Phys. Lett. B \textbf{242}
(1990) 97; Nucl. Phys. B \textbf{366} (1991) 135.
\bibitem{Gorshkov}
V.G.~Gorshkov. Sov. Phys. Usp. \textbf{16} (1973) 322.

\bibitem{semi-hard}
J.D.~Bjorken. FERMILAB-PUB-82-059-THY (1982).

\bibitem{M-fact-scale}
J.M.~Campbell, J.W.~Huston and W.J.~Stirling.  Rep. Prog. Phys.
\textbf{70} (2007) 89; \\
W.~Greiner, S.~Schramm, E.~Stein. \emph{Quantum chromodynamics}.
Springer, Berlin, 2002.

\bibitem{Anastasiou}
Ch.~ Anastasiou, L.~Dixon, K.~Melnikov, and F.~Petriello. Phys. Rev.
D \textbf{69} (2004) 094008.

\bibitem{Owens}
J.F.~Owens. Rev. Mod. Phys. \textbf{59} (1987) 465.

\bibitem{CGC}
F.~Gelis, E.~Iancu, G. Jalillian-Marian, and R.~Venugopalan. Annu.
Rev. Nucl. Part. Sci. (2010).

\bibitem{Barone-Predazzi}
V.~Barone, E.~Predazzi. High energy particle diffraction. Springer,
Berlin, 2002.

\bibitem{dipole-DIS}
V.~Barone, M.~Genovese, N.N.~Nikolaev, E.~Predazzi, and
B.G.~Zakharov. Z. Phys. C \textbf{58} (1993) 541; Phys. Lett. B
\textbf{326} (1994) 161; \\
N.N.~Nikolaev and B.G.~Zakharov. Z. Phys. C \textbf{64} (1994) 631; \\
A.H.~Mueller. Nucl. Phys. B \textbf{415} (1994) 373.

\bibitem{Kimber-Martin-Ryskin}
M.A.~Kimber, A.D.~Martin, and M.G.~Ryskin. Eur. Phys. J. C
\textbf{12} (2000) 655; Phys. Rev. D \textbf{63} (2001) 114027.

\bibitem{geom-scaling}
A.M.~Stasto, K.J.~Golec-Biernat, and J.~Kwiecinski. Phys. Rev. Lett.
\textbf{86} (2001) 596.

\bibitem{DY-resummation}
E.~Fink. arXiv:hep-ph/0105276.

\bibitem{Tevatron-QT}
V.M.~Abazov et al., D0 Collaboration. Phys. Lett. B \textbf{693}
(2010) 522–530.

\bibitem{LHC-QT}
G.~Aad et al., ATLAS Collaboration. arXiv:1107.2381.

\bibitem{W-QT-Tevatron}
V.M. Abazov et al., D0 Collaboration. Phys. Lett. B \textbf{513}
(2001) 292.



\end{thebibliography}
\end{document}